\documentclass[10pt,twocolumn,english,nofootinbib]{revtex4}
\usepackage[T1]{fontenc}
\usepackage[latin9]{inputenc}
\usepackage[a4paper]{geometry}
\usepackage{amsmath}
\usepackage{amssymb}
\usepackage{graphicx}

\makeatletter
\@ifundefined{textcolor}{}
{%
 \definecolor{BLACK}{gray}{0}
 \definecolor{WHITE}{gray}{1}
 \definecolor{RED}{rgb}{1,0,0}
 \definecolor{GREEN}{rgb}{0,1,0}
 \definecolor{BLUE}{rgb}{0,0,1}
 \definecolor{CYAN}{cmyk}{1,0,0,0}
 \definecolor{MAGENTA}{cmyk}{0,1,0,0}
 \definecolor{YELLOW}{cmyk}{0,0,1,0}
}

\def\eq#1{{eq.~(\ref{#1})}}
\def\Eq#1{{Eq.~(\ref{#1})}}

\def\Tr{\mbox{Tr}\,}

\def\hbar{\hspace{0pt}\raisebox{1pt}{$-$} \hspace{-7pt} h}

\def\5{\overline 5}

\newcommand{\be}{\begin{equation}}
\newcommand{\ee}{\end{equation}}
\newcommand{\bea}{\begin{eqnarray}}
\newcommand{\eea}{\end{eqnarray}}
\newcommand{\nn}{\nonumber}

\newcommand{\bwt}{\begin{widetext}}
\newcommand{\ewt}{\end{widetext}}

\makeatother

\usepackage{babel}
\begin{document}


\title{A renormalization group improved  computation of correlation \\functions in
theories
with non--trivial phase diagram
}


\author{Alessandro Codello$^{\diamond}$ }
\author{Alberto Tonero$^{\star}$}
\affiliation{$^{\diamond}$CP3-Origins \& the Danish IAS University of Southern Denmark, Campusvej 55, DK-5230 Odense M, Denmark}
\affiliation{$^{\star}$ICTP-SAIFR \& IFT-UNESP, Rua Dr. Bento Teobaldo Ferraz 271, 01140-070 S\~ao Paulo, Brazil }

\begin{abstract}
We present a simple and consistent way to compute correlation functions
in interacting theories with non--trivial phase diagram.
As an example we show how to consistently compute the four--point function in three dimensional $\mathbb{Z}_2$--scalar theories.
The idea is to perform the path integral by weighting the momentum modes that contribute to it according to their renormalization group (RG) relevance,
i.e. we weight each mode according to the value of the running couplings at that scale.
In this way, we are able encode in a loop computation the information regarding the RG trajectory along which we are integrating.
We show that depending on the initial condition, or initial point in the phase diagram, we obtain different behaviors of the four--point function at the end point of the flow.\\
\\
Preprint: CP3-Origins-2015-006 DNRF90 and DIAS-2015-6
\end{abstract}

\preprint{CP3-Origins-2015-006 DNRF90 and DIAS-2015-6}

\maketitle


\section{Introduction}

One problem that affects computations in quantum field theory (QFT), in both its applications to condensed matter or high energy physics, is the possible appearance of infrared (IR) divergences in the massless or critical cases \cite{Parisi:1988nd}. In this class of theories, the breakdown of the mean field, or Landau, approximation
was one of the original problems that triggered the development of the renormalization group (RG) theory \cite{Wilson}.
This lead to the concept of theory space, running couplings and ultimately the understanding of universality. We then learned how to compute universal quantities like critical exponents or universal ratios but, ultimately, a technique free of IR divergencies that allows the computation of correlation functions, which are the main object of interest in QFT, did not emerge. Only later conformal field theory (CFT) was developed to fill this gap, but only at criticality, i.e. at fixed points \cite{Belavin:1984vu}. Also, applications of CFT techniques have been limited to two dimensions and only recently progress has been made in three dimensions, 
although it has been mostly numerical \cite{ElShowk:2012ht}.

In this paper we want to fill this gap by showing how to compute correlation functions, or more precisely one--particle--irreducible (1PI) vertices (generated by the effective action), along a RG trajectory, in such a way that if this trajectory is critical or massless and hits a fixed point in the IR, no divergencies occur and a perfectly finite analytical result emerges. If instead, the trajectory leads to a massive theory in the IR, then still the computation goes through. This reflects the fact that a theory is indeed a trajectory in theory space that connects a microscopic, or bare action, to the effective action in the IR, which contains the physics.

The idea is to perform the path integral by weighting the momentum modes that contribute to it according to their RG relevance, i.e. we weight each mode according to the value of the running couplings at that scale. In other words, we apply the RG improvement procedure also to the correlation functions by integrating their RG flow with the prior knowledge of the phase space structure, i.e. with the running couplings as known functions of the scale, functions obtained by integrating their respective beta functions. After explaining the method in abstract, we will show how it concretely works in the case of a model with non--trivial phase diagram as the three dimensional $\mathbb{Z}_2$--scalar theory.

\section{The method}

The fundamental reason for the success of RG computations resides in the RG improvement, which usually, and originally, is applied only to the couplings:
here we will apply it to the full effective action. The starting point it the trace--log formula for the one--loop effective action:
\be
\label{1LEA}
\Gamma = S + \frac{1}{2} \textrm{Tr} \log S^{(2)}\,,
\ee
where $S$ is the bare action and $S^{(2)}$ is its Hessian. The trace on the r.h.s. stands (in flat spacetime) for an integration over coordinate space, or momentum space, and for an integration over any representation of internal and spacetime symmetries that the fields may carry. 
Obviously \eq{1LEA} has to be regularized to make sense: let's use a simple mass regularisation $\mu$ for the IR and a cutoff regularisation $\Lambda$ for the ultraviolet (UV).
In this way the effective action will become dependent on $\mu$ through the mass regulator, while the bare action and the trace will implicitly depend on $\Lambda$:
\be
\label{1LEA_2}
\Gamma_\mu = S_\Lambda + \frac{1}{2} \textrm{Tr}_\Lambda \log \Big[ S^{(2)}_\Lambda + \mu^2 \Big]\,,
\ee
Note that all the $\Lambda$ dependence arising from the trace is compensated by counter--terms contained in the bare action $S_\Lambda$ and so the effective action depends only on $\mu$.

Now let's pause a moment to remember how we usually implement the RG in the case of a coupling, say $\lambda$ dimensionless.
We reabsorb the UV divergencies into the renormalized coupling: $\lambda_\mu = \lambda_\Lambda - C \lambda_\Lambda^n  \log \frac{\Lambda}{\mu}$.
To obtain the beta function we then take a derivative with respect to $\mu$ to obtain $\mu \frac{d}{d\mu}\lambda_\mu =  C \lambda_\Lambda^n$, then we perform the RG improvement on the r.h.s. $\lambda_\Lambda \to \lambda_\mu$ to finally obtain the RG equations $\mu \frac{d}{d\mu}\lambda_\mu =  C \lambda_\mu^n$. This is basically the original procedure of Gell-Man and Low \cite{GellMann:1954fq}.

It is at this point that we will deviate from the standard practice and introduce our method by repeating the RG procedure but at the level of \eq{1LEA_2}. For convenience, from here onwards we change notation and we will denote the RG scale with $k$ instead of $\mu$.
By taking a scale derivative of \eq{1LEA_2} we get
\be
\label{ERGE_0}
\partial_t \Gamma_k = \frac{1}{2} \textrm{Tr}_\Lambda \frac{2k^2}{S^{(2)}_\Lambda + k^2}\,,
\ee
where $t=\log k/ k_0$ and $ k_0$ is some reference scale. At this point we perform the RG improvement substituting $S^{(2)}_\Lambda \to S^{(2)}_k$ on the r.h.s. of \eq{ERGE_0}
\be
\label{ERGE_1}
\partial_t \Gamma_k = \frac{1}{2} \textrm{Tr} \frac{ \partial_t R_k}{S^{(2)}_k + R_k}\,,
\ee
where we stepped ahead and introduced the cutoff kernel function $R_k = k^2$. In general, this kind of cutoff is not sufficient to ensure the trace to be finite. 
In order to make \eq{ERGE_1} not UV divergent, $R_k$ is usually taken to be a function of the Laplacian and required to fast fall--off for large values of its argument. In this case also its $t$--derivative decays fast for large argument values and this makes the trace in the r.h.s. of \eq{ERGE_1} convergent. 
In \eq{ERGE_1} the form of $S_k$ is the same of the bare action $S_\Lambda$ with all bare couplings replaced by their running counterparts.

Knowing the flow of $\Gamma_k$, in analogy to the case of the couplings, it is possible to compute the renormalized effective action by integrating \eq{ERGE_1} between the UV and IR
\be
\label{ERGE_2}
\Gamma_0= \Gamma_\Lambda- \frac{1}{2} \int_{0}^\Lambda \frac{dk}{k}\, \textrm{Tr} \frac{ \partial_t R_k}{S^{(2)}_k + R_k}\,,
\ee
where $\Gamma_\Lambda=S_\Lambda$ 
is the bare action which is taken as the initial condition of the flow in the UV and $\Gamma_0\equiv \Gamma_R$ is the renormalized effective action. In order not to create confusion, we want to stress that in this formalism the subscripts "$0$" naturally denotes renormalized quantities and will be often traded for the subscript "$R$".

The action $\Gamma_R$ is the generating functional of 1PI vertices
\be 
\langle \varphi(x_1) \varphi(x_2)\ldots\varphi(x_n)\rangle_{{\tiny \mbox{1PI} }}=\frac{\delta \Gamma_R[\varphi]}{\delta \varphi(x_1)\ldots\delta\varphi(x_n)}\Bigg|_{\varphi=0}\,.
\ee
Finally, one can do an even further step and switch on all possible renormalized couplings, that we imagine to be equal to zero in the bare action, and make the substitution $S_k^{(2)} \to \Gamma_k^{(2)}$ in \eq{ERGE_2}, in this way obtaining a closed equation for the running effective action $\Gamma_k$. One can show that this equation is exact \cite{Wetterich_1993}.

\section{The model}

We start by considering real scalar theories in $d$--dimensional Euclidean space with $\mathbb{Z}_2$--symmetry defined by the following bare action:
\be
S_{\Lambda} = \int d^dx\left\{\frac{1}{2}\partial_\mu\varphi\partial^\mu\varphi+\frac{m^2_\Lambda}{2}\varphi^2+\frac{\lambda_\Lambda}{4!}\varphi^4\right\}\,.
\label{bare}
\ee
We are interested in computing the effective action $\Gamma_0$ using \eq{ERGE_1} and \eq{ERGE_2}, as described in the previous section.
The effective action is a complicated non--local action; here we will perform an expansion in the field $\varphi$ and retain all terms up to order $\varphi^4$.
We thus fix the form of the scale dependent effective action $\Gamma_k$ to be
%
\begin{align}
\Gamma_{k}\!=\!\int \!d^dx\,\Big\{\!\frac{1}{2}\varphi[-\square+\Sigma_{k}(-\square)+m^2_k]\varphi\!+\!\frac{1}{4!}\varphi^2 F_{k}(-\square)\varphi^2\!\Big\}\,,\label{ansatzscalargamma}
\end{align}
%
where $\Sigma_k(-\square)$ is the running polarization function, $m^2_k$ is the running mass of the scalar field and $F_{k}(-\square)$ is the running four--point structure function. Here $\square=\partial_\mu \partial^\mu$ stands for the flat space Laplacian. The constant part of the four--point structure function defines the running quartic scalar coupling $\lambda_k$, i.e. $F_k(0)=\lambda_k$.
The full effective action in principle contains also all terms with a higher field power
but, since they are not relevant for our discussion, we neglect them and restrict just to the truncation of \eq{ansatzscalargamma}.

We now take the Hessian of the bare action in \eq{bare} and make the RG improvement of the couplings: $m_\Lambda \to m_k$ and $\lambda_\Lambda \to \lambda_k$.
We find:
\be 
S_k^{(2)}=-\square+m_k^2+\frac{\lambda_k}{2}\varphi^2\,,
\ee
which is the quantity that has to be inserted in \eq{ERGE_1}.
In order to properly account for threshold effects, it is convenient to choose the argument of the cutoff $R_k(z)$ to be $z=-\square+\frac{\lambda_k}{2}\varphi^2$. In this way \eq{ERGE_1} assumes the form:
\be 
  \partial_t \Gamma_k = \frac{1}{2}\Tr h_{k}\left(-\square+\frac{\lambda_k}{2}\varphi^2,m_k^2\right)\,,\label{scalarfrg}
\ee
where we defined the threshold function
%
$
h_{k}(z,\omega)=\frac{\partial_{t}R_{k}(z )}{z +\omega+R_{k}(z)}\,.
$
%
We need now to evaluate the functional trace in \eq{scalarfrg}; this can be easily done using the non--local heat kernel expansion, which we review in the Appendix.
The use of the expansion reported in \eq{nlhk} leads us to:
\bwt
\begin{eqnarray}
  \partial_t \Gamma_k & = &\frac{1}{2}\int_{0}^{\infty}ds\,\tilde{h}_{k}(s,m_k^{2})\,\textrm{Tr}\, e^{-s\left(X+\frac{\lambda_k}{2}\varphi^2\right)}\nn\\
&=&\frac{1}{2}\frac{1}{(4\pi)^{d/2}}\int d^dx\int_{0}^{\infty}ds\,\tilde{h}_{k}(s,m_{k}^{2})\, s^{-d/2}
\left[1-\frac{\lambda_k}{2}s\varphi^2
+\frac{\lambda_k^2}{4}s^{2}\varphi^2 f_U(sX)\varphi^2 + O(\varphi^6)
\right]\nn\\
& = & \frac{1}{2}\frac{1}{(4\pi)^{d/2}}\int d^dx\left\{ Q_{\frac{d}{2}}\left[h_{k}\right]
-\frac{\lambda_k}{2}Q_{\frac{d}{2}-1}\left[h_{k}\right]\varphi^2
+\frac{\lambda_k^2}{8}\varphi^2\int_0^1 d\xi\, 
Q_{\frac{d}{2}-2}\left[h_k^{X \xi(1-\xi)}\right]
 \varphi^2  + O(\varphi^6) \right\} \,,
 \label{scalarnlhk}
\end{eqnarray}
\ewt
where $X= -\square$. In the last line we have used the explicit expression of $f_U$ given in \eq{sf} and we introduced the $Q$--functional notation, also described in the Appendix.
For computational convenience we have also dropped, on the r.h.s. of equation (\ref{scalarnlhk}), terms arising from the scale derivative $\partial_t R_k$ present in the function $h_k$ and proportional to the beta function of $\lambda_k$.  The neglection of these terms renders the flow equation (\ref{scalarnlhk}) similar to the proper--time RG equation studied, for example, in \cite{Bonanno:2004pq}.
For a detailed discussion about the implications of this approximation see \cite{Litim:2002hj}.

Plugging in the l.h.s. of \eq{scalarnlhk} the ansatz for $\Gamma_k$ given in \eq{ansatzscalargamma} and comparing the same field monomials on both sides, we can read off the flow equations:
\begin{eqnarray}\label{bfls}
\partial_{t}\Sigma_{k}(X) & = & 0
\nonumber\\
\partial_{t}m_{k}^{2} & = & -\frac{1}{2}\frac{\lambda_{k}}{(4\pi)^{d/2}}Q_{\frac{d}{2}-1}\left[h_{k}\right]
\nonumber\\
\partial_{t}F_{k}(X)& =&\frac{3}{2}\frac{\lambda_{k}^{2}}{(4\pi)^{d/2}}\int_{0}^{1}d\xi\, Q_{\frac{d}{2}-2}\left[h_{k}^{X\xi(1-\xi)}\right]\,,
\end{eqnarray}
where we defined $h_k^a(z,\omega) \equiv h_k(z+a,\omega)$.
The last line of \eq{bfls} reduces to the flow equation of the quartic coupling $\lambda_k$ in the limit $X\to 0$:
\begin{equation}
\partial_{t}\lambda_{k}  =  \frac{3}{2}\frac{\lambda_{k}^{2}}{(4\pi)^{d/2}}Q_{\frac{d}{2}-2}\left[h_{k}\right]\,.
\label{lambda}
\end{equation}
The four--point structure function is computed integrating the flow of $F_k(X)$ in \eq{bfls} from $k=\Lambda$ (UV) down to $k=0$ (IR):
\bea \label{4psf}
\Delta F(X)&=&\int_0^\Lambda \frac{dk}{k}\,\partial_{t}F_{k}(X) \nn \\
&=&\frac{3}{2}\frac{1}{(4\pi)^{d/2}}\int_0^\Lambda \frac{dk}{k}\lambda_{k}^{2}\int_{0}^{1}d\xi\, Q_{\frac{d}{2}-2}\left[h_{k}^{X\xi(1-\xi)}\right]\,.\nn\\
\eea
The function $F_{0}(X)=F_{\Lambda}(X)-\Delta F(X)$, where $F_{\Lambda}(X)=\lambda_\Lambda$ is the UV initial condition, 
completely determines the 1PI four--point correlation function:
%
\bea 
&&\langle \varphi(x_1) \varphi(x_2)\varphi(x_3)\varphi(x_4)\rangle_{\mbox{{\tiny 1PI}}} \qquad\qquad\nn\\
&&\qquad=\frac{1}{6}\delta_{xx_1}\delta_{xx_2}F_{0}(-\square)\delta_{xx_3}\delta_{xx_4}+\nn\;\mbox{permutations}\,;
\eea
%
connected correlations can then be reconstructed via standard relations.

%
\Eq{4psf} is our main result so far, and before moving on to consider an explicit application,
we will stop a moment to explain its meaning and in particular how it is related to the procedure we announced in the introduction.
Consider the massless case $m_k=0$ for simplicity and
let's undo the RG improvement of the self--interaction coupling for a moment by restoring $\lambda_\Lambda$.
Using the fact that $h_k^a = \partial_t \log (z +a+ R_k) \equiv \partial_t l^a_k$ we can rewrite \eq{4psf} as:
\bea 
&&\Delta F(X)= \int_0^\Lambda \frac{dk}{k} \lambda_{\Lambda}^{2}  \,\partial_t \Big\{ \textrm{one--loop} \Big\}
\eea
where 
\bea
\textrm{one--loop} = \frac{3}{2} \frac{1}{(4\pi)^{d/2}} \int_{0}^{1}d\xi\, Q_{\frac{d}{2}-2}\left[l_{k}^{X\xi(1-\xi)}\right]\,,
\label{1loop}
\eea
is the standard one--loop result in presence of the cutoff $R_k$, as the reader can easily convince itself.
Now we can explain the crucial point of our construction: in \eq{1loop} one can perform the RG improvement $\lambda_\Lambda \to \lambda_k$ {\it after} exchanging the $k$--integral with the bare coupling, or {\it before} and then being unable to do the switch. In the first case we simply reproduce the standard one--loop computation since formally $\int_0^\infty \frac{dk}{k} \partial_t = 1$, while in the second case we perform an RG improved computation, here of the structure function $F_0(X)$, where we effectively weight each momentum shell $dk$ with the value of the running coupling at that point, here $\lambda_k^2$.
This second procedure is the one we propose and its application is encoded in \eq{4psf}.
As we will now show with an explicit example, this procedure is more powerful, consistent and well defined than the standard perturbative approach since it automatically encodes the information about of the phase space trajectory we are integrating along, through the scale dependent couplings, here $\lambda_k$ and $m_k$ for which we have previously solved.

\section{The phase diagram}
%
\begin{figure}[ht!]
\begin{center}
\includegraphics[width=3.7in]{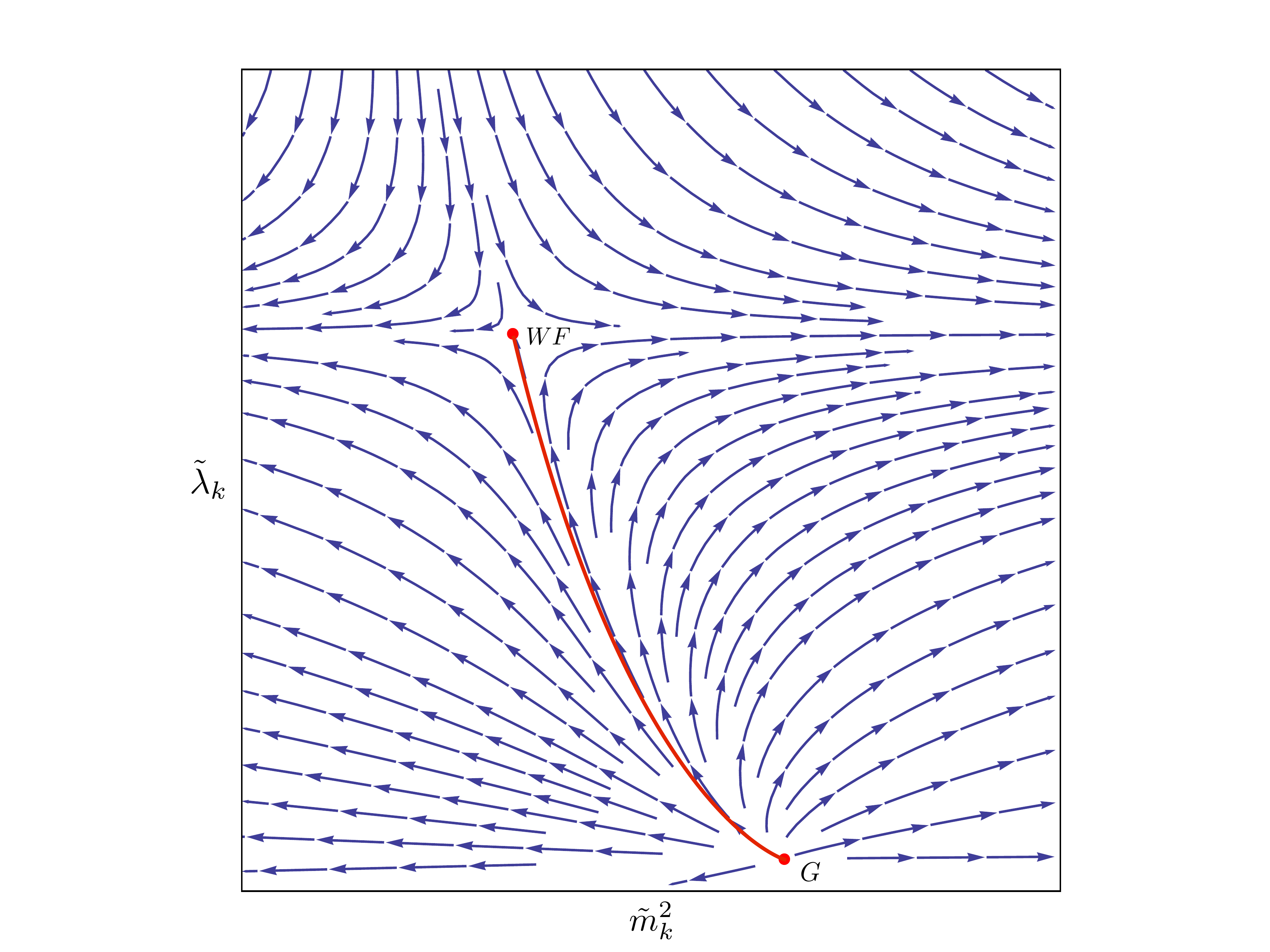}
\caption{Phase diagram in the three dimensional case. One can see the Gaussian and Wilson--Fisher fixed points and the trajectory connecting them, along which we integrate the flow of the structure function $F_k(x)$.}
\end{center}
\end{figure}
%
We restrict ourselves to scalar theories in three dimension, which are UV finite but suffer of IR divergencies in the massless--critical case and are perfect to show how our method works. The beta functions for $m_k^2$ and $\lambda_k$ can be read off by setting $d=3$ in \eq{bfls} and \eq{lambda}:
\bea \label{3Dbf}
\partial_{t}m^2_{k}&=&-\frac{1}{2}\frac{\lambda_{k}}{(4\pi)^{\frac{3}{2}}}Q_{\frac{1}{2}}\left[h_{k}\right]=
-\frac{\lambda_{k}}{4\pi^2}\frac{k^3}{k^2+m^2_k}\nn\\
\partial_{t}\lambda_{k}  &=&  \frac{3}{2}\frac{\lambda_{k}^{2}}{(4\pi)^{\frac{3}{2}}}Q_{-\frac{1}{2}}\left[h_{k}\right]=\frac{3 \lambda_{k}^2}{8\pi^2}\frac{k}{k^2+ m_k^2}\,.
\eea
We have used the explicit values of the $Q$--functionals given in \eq{AQ1half} and \eq{AQm1half} of the Appendix.
The beta function for the dimensionless couplings $\tilde m_k^2=m_k^2/k^2$ and $\tilde \lambda_k=\lambda_k/k$ are:
\bea \label{3Ddbf}
\partial_{t}\tilde m^2_{k}&=&-2 \tilde m^2_{k}
-\frac{\tilde\lambda_{k}}{4\pi^2}\frac{1}{1+\tilde m^2_k}\nn\\
\partial_t \tilde \lambda_{k}&=&-\tilde \lambda_{k} +\frac{3\tilde \lambda_{k}^2}{8\pi^2}\frac{1}{1+\tilde m_k^2}\,.
\eea
There exist two fixed points at:
\be \label{3Dfp}
\begin{array}{l}
\tilde m^2_*=-\frac{1}{3}\qquad \tilde \lambda_*=\frac{16\pi^2}{9}\qquad\mbox{(IR)}\nn\\
\nn\\
\tilde m^2_*=0\qquad\qquad \tilde \lambda_*=0\qquad\mbox{(UV)}
\end{array}
\ee
The IR fixed point is the well--known Wilson--Fisher fixed point and has one attractive and one repulsive direction.
The UV fixed point is the Gaussian one and has both directions attractive, as shown in Fig.\,1.
The critical exponents at the IR fixed point can be easily computed, see for example \cite{Codello:2012ec}.
It is possible to find an analytic solution for the flow if one considers \eq{3Ddbf} in the approximation in which the mass in the denominator on the r.h.s. is neglected:
\bea \label{3Ddbf2}
\partial_{t}\tilde m^2_{k}&=&-2 \tilde m^2_{k}
-\frac{\tilde\lambda_{k}}{4\pi^2}\nn\\
\partial_t \tilde \lambda_{k}&=&-\tilde \lambda_{k} +\frac{3\tilde \lambda_{k}^2}{8\pi^2}\,.
\eea
The values of the two fixed points are slightly shifted:
\be
\begin{array}{l}
\tilde m^2_*=-\frac{1}{3}\qquad \tilde \lambda_*=\frac{8\pi^2}{3}\qquad\mbox{(IR)}\nn\\
\nn\\
\tilde m^2_*=0\qquad\qquad \tilde \lambda_*=0\qquad\mbox{(UV)}
\end{array}
\ee
but their properties are not.
The analytic solution of the system in \eq{3Ddbf2} is:
\bea \label{sol3Ddbf2}
\tilde m^2_{k}&=&\frac{C_2}{k^2}-\frac{2C_1}{k}+\frac{6C_1^2}{k^2}\log(3C_1+k)
\nn\\
\tilde \lambda_{k}&=&\frac{8C_1\pi^2}{3C_1+k}\,,
\eea
where $C_1$ and $C_2$ are integration constants with, respectively, dimension of mass and mass squared. Notice that the UV fixed point is reached, for $k\to \infty$, independently of the values of these constant, since it is attractive in both directions. On the other hand, the IR fixed point is reached, for $k\to 0$, only if a particular relation between the integration constants holds, i.e. only if we are on the critical trajectory. 

We want to repeat here the fundamental point: a theory is represented by a trajectory in theory space.
All trajectories can be parametrized by the value of the renormalized mass
\be \label{renm}
m^2_R=\lim_{k\to 0} k^2 \tilde m_k^2=C_2+6C_1^2\log(3C_1)\,.
\ee
The unique trajectory that flows between the two fixed points is defined by:
\be \label{finrel}
m^2_R=0\,.
\ee
Using this condition in \eq{sol3Ddbf2} gives the following solutions:
\bea \label{sol3Ddbf2b}
\tilde m^2_{k}&=&-\frac{2C_1}{k}+\frac{6C_1^2}{k^2}\log\left(\frac{3C_1+k}{3C_1}\right)
\nn\\
\tilde \lambda_{k}&=&\frac{8C_1\pi^2}{3C_1+k}\,.
\eea
\Eq{sol3Ddbf2b} describes the unique trajectory in the $(\tilde m^2_k,\tilde\lambda_k)$ plane that flows from the Gaussian fixed point in the UV to the Wilson--Fisher fixed point in the IR. In Fig.\,1 this finite trajectory is shown in red. 
All non critical trajectories are characterized by the condition $m^2_R\neq 0$. Positive values of $m^2_R$ label trajectories that flow towards the symmetric phase, while negative values of $m^2_R$ label trajectories that flow towards the broken phase.

We will now solve the theory along the critical line connecting the Gaussian and Wilson--Fisher fixed points and compare the result with the standard perturbative treatment. 

\section{The four--point function}

The four--point structure function is obtained by evaluating the expression in \eq{4psf} for $d=3$.
We have
\be \label{df3D}
\Delta F(X)=\frac{3 }{8\pi^2}\frac{1}{\sqrt{X}}\int_0^\infty\frac{dk}{k}\lambda_{k}^2\frac{k^2}{k^2+m^2_k} g\!\left(\frac{X}{k^2}\right)\,,
\ee
where we have used the explicit form of the $Q$--functional given in \eq{AintQm1half} to compute
\be
g(u)=\log\frac{2+\sqrt{u}}{2-\sqrt{u}}\theta(2-\sqrt{u})+\log\frac{\sqrt{u}+2}{\sqrt{u}-2}\theta(\sqrt{u}-2)\,.
\ee
This is the analytic result we obtained by using the optimized cut-off in \eq{optcut}. Other cutoffs may be implemented and in general will give different forms for $g(u)$, i.e. $g(u)$ is a regulator dependent quantity.
We also removed the UV cutoff $\Lambda \to \infty$ since the theory is finite in this limit.
%
\begin{figure}[ht!]
\begin{center}
\includegraphics[width=3.3in]{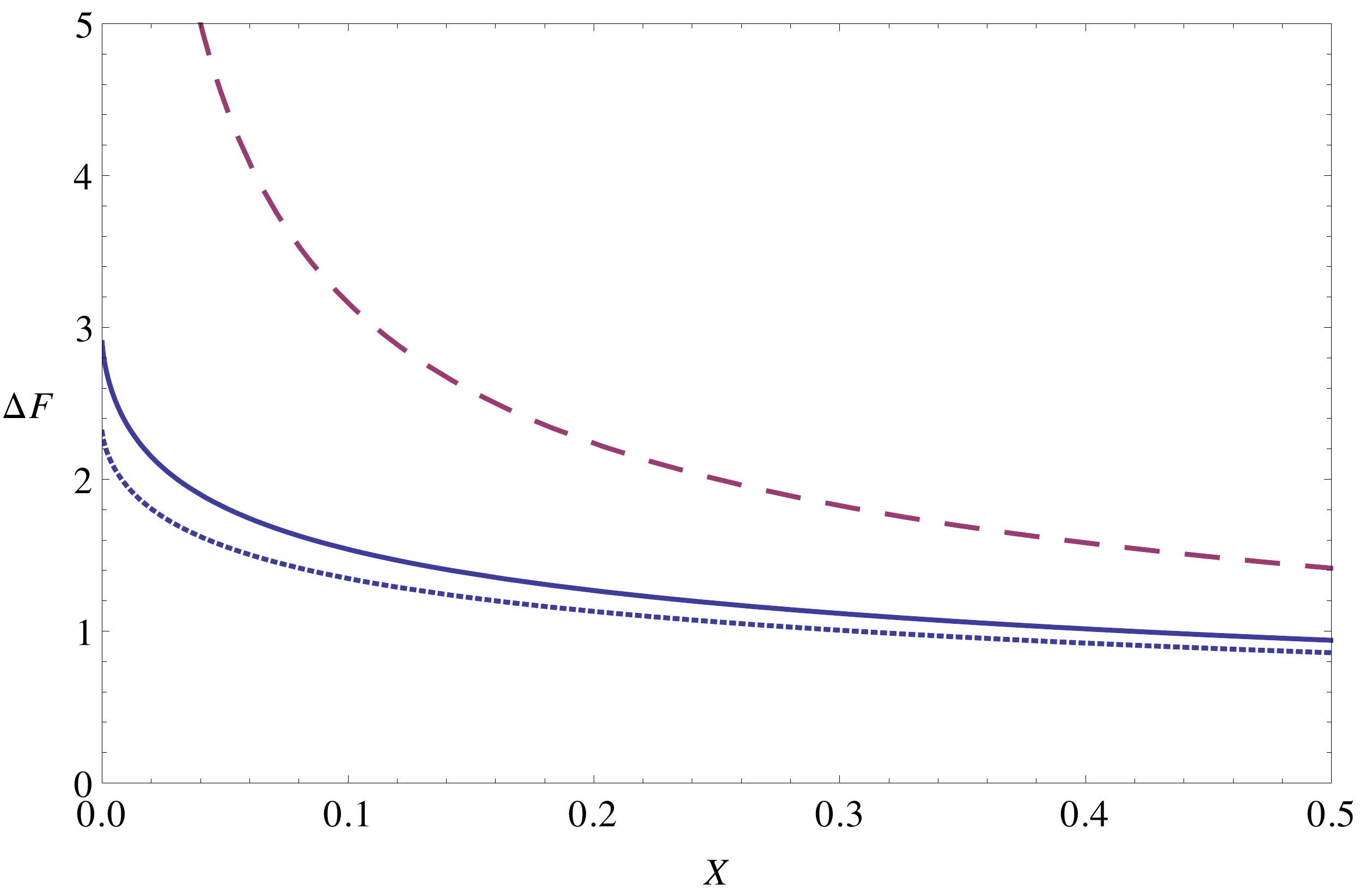}
\caption{The dashed curve is the standard perturbative result $\Delta F^{1L}(X,\lambda_R,0)$ in the massless limit, the dotted curve represents the improved structure function $\Delta F^I(X,{\lambda_R}/{8\pi^2})$ in the $m_k=0$ case and the solid curve represents the full improved structure function $\Delta F^{II}(X,{\lambda_R}/{8\pi^2},0)$. All the three structure functions are evaluated for $\lambda_R=4/\sqrt{3}$.}
\end{center}
\end{figure}

\subsection{One--loop computation}

The structure function at one--loop is obtained by performing the RG improvement only
after having exchanged the bare coupling and the scale integral in \eq{df3D} and by setting the mass to its renormalized value:
\bea \label{df3D1la}
\Delta F^{1L}(X,\lambda_R,m^2_R)&=&\frac{3 \lambda^2_R}{8\pi^2}\frac{1}{\sqrt{X}}\int_0^\infty\frac{dk}{k}\frac{k^2}{k^2+m^2_R} g\!\left(\frac{X}{k^2} \right)\nn\\
&=&\frac{3\lambda^2_R}{8\pi \sqrt{X}}\mbox{Arctan}\frac{\sqrt{X}}{2m_R}\,,
\eea
where $m_R$ and $\lambda_R$ are the renormalized couplings.
In the massless limit \eq{df3D1la} reduces to:
\be \label{df3D1lb}
\Delta F^{1L}(X,\lambda_R,0)=\frac{3 \lambda_R^2}{16}\frac{1}{\sqrt{X}}\,.
\ee
\Eq{df3D1la} and \eq{df3D1lb} are known to be the correct perturbative results~\cite{Parisi:1988nd,ZinnJustin:2002ru}.
Notice that in the perturbative case the result is not sensitive the the RG trajectory we are on, i.e. there is no way to codify this information and the result is the same in all cases. In particular, the massless (critical) expression in \eq{df3D1lb} is IR divergent at large scales $X \to 0$, signalling the breakdown of Landau--mean field theory.
In general, for $d<4$, the IR divergences at $m_R=0$ are more and more important as the perturbative order increases.
If $p=\sqrt{X}$ represents the momentum scale, the four--point function, computed at one--loop through a "bubble" diagram, behaves as $p^{-(4-d)}$ and its iteration $n$ times inside UV convergent diagrams, will produce IR divergencies if $n(4-d)>d$. For more details see~\cite{Parisi:1988nd}. We will see in the next section that the RG improved computation of the four--point function structure function (bubble diagram) will cure this problem.
\\
\\
\\
\\

\subsection{RG improved computation}

We now perform the RG improved calculation where the scale integration contains the running couplings $m_k$ and $\lambda_k$ evaluated on the critical trajectory connecting the two fixed points. Every momentum shell will thus be weighted accordingly to the strength of the interaction and, differently from the perturbative case, we will obtain an IR finite four--point structure function.

For a better analytical understanding, we will first integrate the flow of the structure function \eq{df3D} in the case $m_k=0$, which we will call $\Delta F^I$. In this case theory space is one dimensional and the complete $k$--dependence of $\lambda_k$ is given in \eq{sol3Ddbf2b}. We then find the following result:
\bwt
\bea \label{df3Dfnm}
\Delta F^I(X, C_1)&=&\frac{ 24 C_1^2 \pi^2}{\sqrt{X}}\int_0^\infty dk\,\frac{k}{(3C_1+ k)^2}g\!\left(\frac{X}{k^2} \right)=\frac{6C_1^2\pi^2}{\sqrt{X}}\Bigg\{\pi^2+\frac{48\,C_1\sqrt{X}}{X-36 \,C_1^2}\log \frac{6C_1}{\sqrt{X}}+2\left(\log\frac{\sqrt{X}+6C_1}{\sqrt{X}-6C_1}\right)^2\nn\\&&-2\left(\log\frac{6C_1+\sqrt{X}}{6C_1-\sqrt{X}}\right)^2
-4  \mbox{PolyLog}\!\left[2,\frac{\sqrt{X}+6C_1}{\sqrt{X}-6C_1}\right]+4  \mbox{PolyLog}\!\left[2,\frac{6C_1+\sqrt{X}}{6C_1-\sqrt{X}}\right]
\Bigg\}\,.
\eea
\ewt
The value of the integration constant $C_1$ can be fixed by imposing that $\Delta F^I(X,C_1)$
has to reproduce the perturbative result  in the UV limit $X\to \infty$:
\be 
\Delta F^I(X,C_1)\Big|_{X \to \infty}\sim \frac{12 \pi^4 C_1^2}{\sqrt{X}}\,.
\ee
Comparing with \eq{df3D1lb} we then have:
\be 
12 \pi^4 C_1^2=\frac{3 \lambda_R^2}{16}\qquad\Rightarrow \qquad C_1=\frac{\lambda_R}{8\pi^2}\,.
\ee
In the IR limit $X\to 0$ the improved structure function is given by:
\be 
\lim_{X \to 0} \Delta F^I(X,{\lambda_R}/{8\pi^2})= \lambda_R
\ee
and is finite as should be.
The improved structure function $\Delta F^I(X, {\lambda_R}/{8\pi^2})$ differs from the standard one--loop perturbative result $\Delta F^{1L}(X,\lambda_R,0)$ of \eq{df3D1lb} at small values of $X$. A comparison between the perturbative and the improved computation is shown in Fig.\,2, where we have set $\lambda_R=4/\sqrt{3}$.
The improved result takes correctly into account the fact that the theory flows to an IR fixed point and thus it furnishes a IR finite quantity.
In the language of perturbation theory this is equivalent to say that the bubble diagram is finite and no more divergent as $1/\sqrt{X}$ in the IR.
This is the exemplification of how the RG cures the breakdown of Landau--mean field theory if properly implemented, as does the method we propose.

We now relax the condition $m_k=0$ and
we integrate the flow of the structure function in \eq{df3D} substituting for $m^2_k$ and $\lambda_k$
their complete $k$--dependence as given by \eq{sol3Ddbf2}. We have:
\be \label{df3Df}
\Delta F^{II}(X,C_1,m_R^2)=\frac{3 }{8\pi^2}\frac{1}{\sqrt{X}}\int_0^\infty dk\, k\, \frac{\tilde \lambda_{k}^2}{1+\tilde m^2_k}g\!\left(\frac{X}{k^2} \right)\,,
\ee
where $C_2$ has been traded for $m_R^2$ using the relation in \eq{renm}.
We first perform a numerical integration of \eq{df3Df} along the critical trajectory ($m_R^2=0$) setting $C_1=\lambda_R/8\pi^2=1/\sqrt{12}\pi^2$ in order to compare the result directly with the previous computations, as shown in Fig.\,2. Again the four--point structure function is finite in the IR, namely for $X\to 0$.
The integration of \eq{df3Df} along non critical trajectories with $m_R^2>0$ provide also a finite IR limit of the structure function. Fig.\,3 shows the behavior of the structure function obtained by numerical integration of \eq{df3Df} considering different positive values of $m^2_R$ and setting $C_1=1/\sqrt{12}\pi^2$.
Trajectories that flow to the broken phase ($m_R^2 <0$) make the integrand of \eq{df3Df} to develop a singularity along the integration path and we can not describe this situation properly within our truncation. In this case the truncation we use turns out to be inconsistent because it describes a theory which is quantized around the wrong vacuum.
At the fixed point the theory is a CFT, so the $X\to0$ limit can be compared with what is known from CFT.
Consistently we find that the 1PI part of the four--point function, i.e. with the four propagator stripped--off, is to first approximation constant~\cite{Dolan:2000ut}.

\begin{figure}[ht!]
\begin{center}
\includegraphics[width=3.3in]{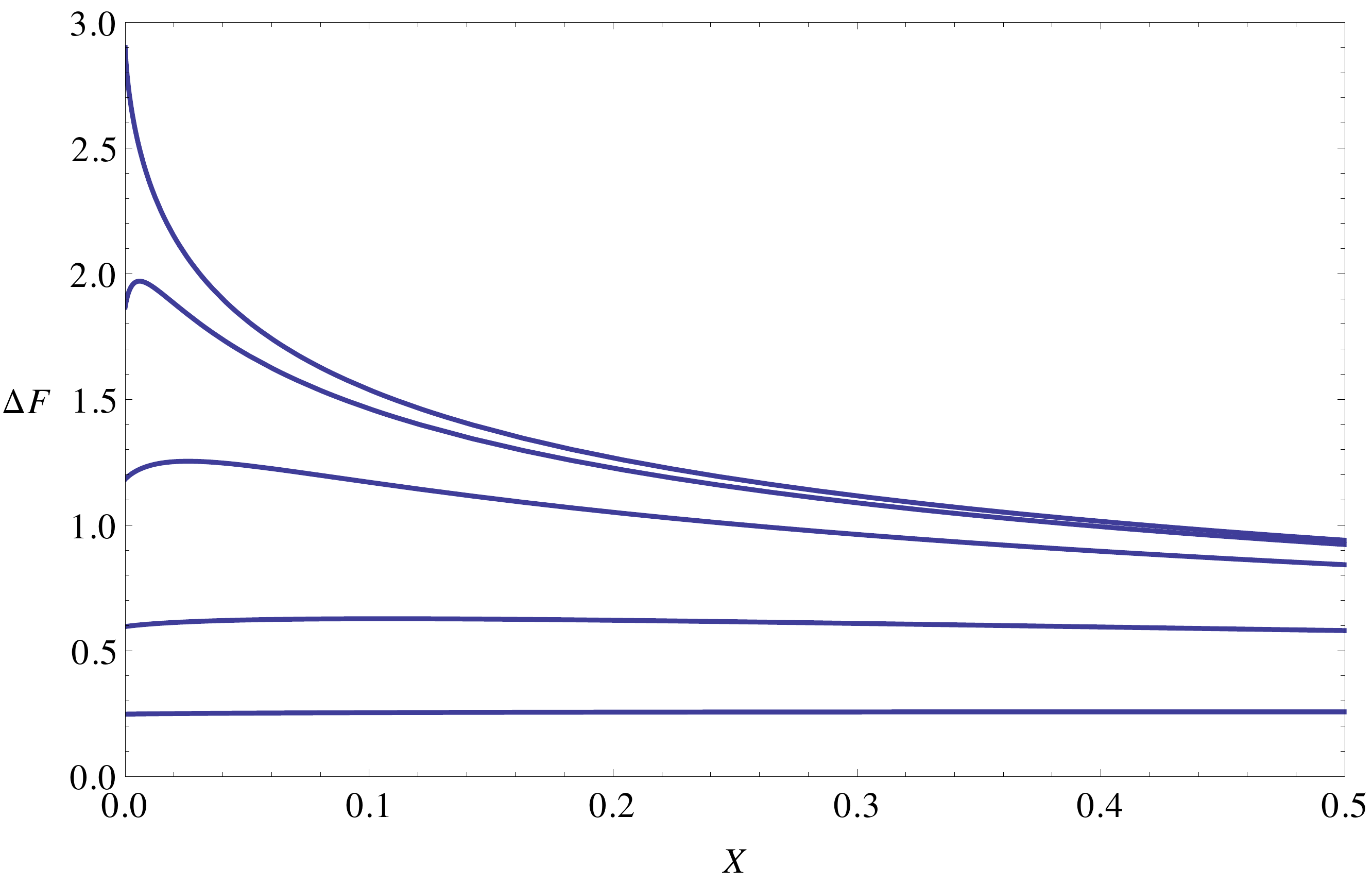}
\caption{The plot represents different behaviors of the structure function $\Delta F^{II}(X,1/\sqrt{12}\pi^2,m_R^2)$ 
computed, starting from below, for $m_R^2=1$, $0.1$, $0.01$, $0.001$ and $0$.}
\end{center}
\end{figure}
\section{Outlook}

In this paper we showed that it is possible to extend the standard RG improvement of the couplings to an improvement of the full effective action by generalising the original procedure Gell--Mann and Low. This method can be used to make any standard one--loop computation include the non--trivial information encoded in the phase diagram of the theory.
In this way it is possible to cure both UV and IR divergencies when appropriate fixed points are available as, respectively, starting or ending point of the RG flow.
We also showed how this makes one--loop computations become sensitive to the RG trajectory along which one integrates the theory, as it should.

We exemplified the method with an application to a three dimensional real scalar with  $\mathbb{Z}_2$--symmetry and showed how to obtain a consistent IR finite four--point structure function as the result of the integration along the trajectory connecting the Gaussian (UV) fixed point with the Wilson--Fisher (IR) fixed point.
 
Our method can be applied to any QFT, for example to improve a mean field analysis.
It also works the other way around, i.e. going to the UV; in this case one can obtain finite results in theories with an UV fixed point.
For example, in four dimensions one can apply these ideas to the computation of the gravitational effective action along a trajectory controlled by an UV fixed point;
a first step in this direction has been done in \cite{Satz:2010uu}.
Other applications can be to NL$\sigma$M \cite{Codello:2008qq} or to controllable asymptotically safe theories \cite{Litim:2014uca}.

\begin{acknowledgments}
The work of AC is supported by the Danish National Research
Foundation under the grant number DNRF:90.
The work of AT was supported by the S\~ao Paulo Research Foundation (FAPESP) under grants 2011/11973-4 and 2013/02404-1.
\\
\end{acknowledgments}


\appendix

\section{non--local heat kernel}

We now show how to compute the functional trace in \eq{scalarfrg}:
\be
\label{wetterich}
\partial_t \Gamma_k =\frac{1}{2}\Tr h_k(-\square +U,\omega)\,,
\ee
with $U\equiv\frac{\lambda_k}{2}\varphi^2$.
Introducing the Laplace transform $\tilde h_k(s,\omega)$,
\be
h_k(z,\omega)=\int_0^\infty ds\, \tilde h_k(s,\omega) e^{-s z} \,,
\ee
we can rewrite \eq{wetterich} as:
\be
\label{elsa}
\partial_t \Gamma_k=
\frac{1}{2}\int_0^\infty ds\, \tilde h_k(s,\omega) \Tr e^{-s (-\square +U)} \ .
\ee
The trace in the r.h.s of \eq{elsa} is computed using the non--local heat kernel expansion \cite{Codello:2012kq} which retains the infinite number of heat kernel coefficients in terms of non--local structure functions or form--factors and reads:
\bwt
\begin{eqnarray}
\Tr e^{-s(-\square +U)} & = & \frac{1}{(4\pi s)^{d/2}}
\int d^{d}x\sqrt{g}\,\bigg\lbrace \mathbf{1}-s U+Uf_{U}(-s\square)U+O(U^{3})
\bigg\rbrace \,,
\label{nlhk}
\end{eqnarray}
\ewt
where the structure function $f_{U}(x)$ is given by:
\be
f_{U}(x) =  \frac{1}{2}\int_{0}^{1}d\xi\, e^{-x\xi(1-\xi)}\,.\label{sf}
\ee
%

\section{$Q$--functionals}
The $Q$-functionals are defined as
\be \label{deQn}
Q_{n}[h_k^a]=\int_0^\infty\,ds s^{-n}e^{-sa}\tilde h_k(s,\omega)\,,
\ee
where $h_k^a(z,\omega)\equiv h_k(z+a,\omega)$.
For a positive real number $n>0$  it is possible to rewrite the $Q$-functional as a Mellin transform~\cite{Codello:2008vh}:
\be \label{AQn}
Q_{n}[h_k]=\frac{1}{\Gamma(n)}\int_{0}^{\infty}dz\, z^{n-1}h_k(z,\omega)\,,
\ee
while for a negative half--integer $n=-\frac{2m+1}{2}<0$ we have \cite{Codello:2008vh}:
\be 
Q_{-\frac{2m+1}{2}}[h_k]=\frac{(-1)^{m+1}}{\sqrt{\pi}}\int_{0}^{\infty}dz\, z^{-1/2}\frac{d^{m+1}}{dz^{m+1}} h_k(z,\omega)\,.
\ee
We compute the $Q$-functional using an optimized cutoff of the form
\be \label{optcut}
R_k(z)=(k^2-z)\theta(k^2-z)\,.
\ee
In this case the function $h_k(z,\omega)$ takes the following form,
\be 
h_k(z,\omega)=\frac{2k^2\theta(k^2-z)}{z+\omega+(k^2-z)\theta(k^2-z)}\,.
\ee
We have
\bea
Q_{\frac{1}{2}}\left[h_{k}\right]&=&\frac{1}{\Gamma(\frac{1}{2})}\int_{0}^{\infty}dz\, z^{-\frac{1}{2}}h_k(z,\omega)\nn\\
&=&\frac{4}{\sqrt{\pi}}\frac{k^2}{k^2+\omega}\sqrt{k^2}\,, \label{AQ1half}
\eea
\bea
Q_{-\frac{1}{2}}\left[h_{k}^a\right]&=&-\frac{1}{\sqrt{\pi}}\int_{0}^{\infty}dz\, z^{-\frac{1}{2}}\frac{d}{dz}h_k(z+a,\omega)\nn\\
&=&\frac{2}{\sqrt{\pi}} \frac{k^2}{k^2+\omega}\frac{\theta(k^2-a)}{\sqrt{k^2-a}}
\label{AQm1half}
\eea
and
\bwt	
\bea
\int_0^1 d\xi\, Q_{-\frac{1}{2}}\left[h_{k}^{X\xi(1-\xi)}\right]&=&\frac{2}{\sqrt{\pi}} \frac{k^2}{k^2+m_k^2}\int_0^1 d\xi\,\frac{\theta(k^2-X\xi(1-\xi))}{\sqrt{k^2-X\xi(1-\xi)}}\nn\\
&=&\frac{2}{\sqrt{\pi}} \frac{k^2}{k^2+m_k^2}\frac{1}{\sqrt{X}}\left[\log\frac{2k+\sqrt{X}}{2k-\sqrt{X}}\theta(4k^2-X)+\log\frac{\sqrt{X}+2k}{\sqrt{X}-2k}\theta(X-4k^2)\right]\label{AintQm1half}\,.
\eea
\ewt
This completes the evaluation of the $Q$--functionals.

\end{document}